\documentclass[a4paper,11pt]{article}
\usepackage[numbers,sort&compress]{natbib}
\usepackage[all]{xy}
\usepackage{amsmath,amsfonts,amssymb,amsthm,epsfig,amscd,comment,latexsym,psfrag}
\usepackage{CJK}
\usepackage{mathrsfs}
\usepackage{nicefrac,xspace,tikz}
\usepackage{arydshln}
\usetikzlibrary{arrows}
\usepackage{graphicx}
\usepackage{caption,subcaption}
\usepackage{xcolor,float}
\usepackage{appendix}
\usepackage{graphicx}
\usepackage{caption}
\makeatletter

\newcommand{\Rmnum}[1]{\expandafter\@slowromancap\romannumeral #1@}
\makeatother

\def\footnoterule{\kern 1mm \hrule width 7cm \kern 2.2mm}%

\def\dprod{\displaystyle\prod}
\def\dlim{\displaystyle\lim}
\def\dsum{\displaystyle\sum}

\def\tr{\mathrm{tr}}

\allowdisplaybreaks
\begin{document}
\begin{center}
{\Large\bf Large $N$ limit of complex multi-matrix model}\vskip .2in
{\large Lu-Yao Wang$^{a}$,\footnote{wangly100@outlook.com} Yu-Sen Zhu$^{a}$,\footnote{
zhuyusen@cnu.edu.cn} Shao-Kui Yao$^{b}$\footnote{yaoshaokui123@126.com}
Bei Kang$^{c}$\footnote{Corresponding author:kangbei@ncwu.edu.cn} } \vskip .2in
$^a${\small{ School of Mathematical Sciences, Capital Normal University, Beijing 100048, China}} \\
$^b${\small{ School of Mathematical Sciences, Henan Institute of Science and Technology, Xinxiang 453003, Henan, China}} \\
$^c${\small{ School of Mathematics and Statistics, North China University of Water Resources and
Electric Power, Zhengzhou 450046, Henan, China}}\\

\vskip .2in

\begin{abstract}
We construct the complex multi-matrix model with $W$-representation and calculate the correlators.
We establish the correspondence between the connected correlators and length-$2n$ $q$-colored
Dyck walks in Fredkin spin chain and discuss the entanglement entropy. Moreover, we analyze the
free energy of this multi-matrix model. For the leading coefficient of the free energy, it relates to the connected correlators in large $N$ limit.
\end{abstract}

\end{center}



\section{Introduction}
Multi-matrix models were initially developed to relate to minimal matter coupled to quantum gravity
interacting with $c<1$ matter  \cite{M.R. Douglas1990}.
To further study the integrability and $W$ constraints of multi-matrix model, the conformal
multi-matrix models were derived from the conformal field theories \cite{Mironov1991,Mironov1993}.
In addition, as a special case of the multi-matrix model, the partition functions of the
multi-matrix chain models were proposed to describe the constrained hierarchy
\cite{J.Boer1991,H.ARATYN1997}. The one-point functions and constraint equations at finite $N$ of
such models were also derived, and the generating operators of the one-point functions satisfy the
$W_{1+\infty}$ like algebra \cite{C.Ahn1992}. Moreover, the integrability and topological content of
the multi-matrix chain models were analyzed \cite{L.Bonora1995}.
Based on the results of the multi-trace multi-matrix operators \cite{TW0711}-\cite{YK0910}, the correlation functions of Gaussian complex multi-matrix models can be translated into correlation functions of the two-dimensional field theories \cite{Kimura2014}.
The multiple scaling limits and the double scaling limits of the multi-matrix model have been investigated \cite{multiplescaling,Doublescaling}.

$W$-representation provides a dual formula for partition function through differentiation \cite{
Morozov0902}. It is useful to analyze the structures of matrix model, such as the spectral curve
\cite{Mironov2210} and superintegrability \cite{Mironov2201}-\cite{Alexandrov2022}.
The $W$-representations of two-matrix models have been studied \cite{Mironov230104107}-\cite{2matrix}.
The superintegrable multi-matrix model was constructed in Ref.\cite{1405}. It possesses $W$-representation and relates to the hypergeometric Hurwitz $\tau$-function. In this paper, we will construct a new complex multi-matrix model with $W$-representation and analyze the free energy in the large $N$ limit.

\section{Complex multi-matrix model}

\subsection{$W$-representation of complex multi-matrix model}
We construct the  complex multi-matrix model
\begin{eqnarray}\label{partition-C1}
&&Z_{C}\{t\}=\int d^2M_1d^2M_2\cdots d^2M_p \exp[-\sum_{i=1}^p \tr M_iM_i^{\dag}
+\sum_{i=1}^{p}\sum_{k=0}^{\infty}N^{-k}t_k^{(i)}\tr (M_iM_i^{\dag})^{k}
\nonumber\\
&&+\sum_{u=2}^{\infty}\sum_{i_1,\cdots,i_u=1,\atop i_a\neq i_{a+1}}^{p}\sum_{k_1,\cdots
k_{u}=1}^{\infty}N^{-(k_1+\cdots+k_u)}t_{k_1,\cdots ,k_{u}}^{(i_1,\cdots,i_u)}
\tr
(M_{i_1}M_{i_1}^{\dag})^{k_1}(M_{i_2}M_{i_2}^{\dag})^{k_2}\cdots
(M_{i_u}M_{i_u}^{\dag})^{k_u}],
\end{eqnarray}
where $a=1,\cdots u-1$, and the integrands are all~$N\times N$ complex matrices.
Since the trace of the matrix product is symmetric under cycle permutation $\sigma=(i_1,\cdots,i_u)$, i.e,
\begin{eqnarray*}
&&\tr(M_{\sigma(i_1)}M_{\sigma(i_1)}^{\dag})^{k_1}
(M_{\sigma(i_2)}M_{\sigma(i_2)}^{\dag})^{k_2}\cdots(M_{\sigma(i_u)}M_{\sigma(i_u)}^{\dag})^{k_u}
\nonumber\\&&=\tr(M_{i_1}M_{i_1}^{\dag})^{k_1}(M_{i_2}M_{i_2}^{\dag})^{k_2}\cdots
(M_{i_u}M_{i_u}^{\dag})^{k_u},
\end{eqnarray*}
we only consider $\tr(M_{i_1}M_{i_1}^{\dag})^{k_1}(M_{i_2}M_{i_2}^{\dag})^{k_2}\cdots
(M_{i_u}M_{i_u}^{\dag})^{k_u}$ in (\ref{partition-C1}). Here, we do not consider other terms under
the permutation $\sigma$ in (\ref{partition-C1}). In addition, for the $k$-transposition
$(\underbrace{i_1,\cdots,i_s}_{1}; \underbrace{i_1,\cdots,i_s}_{2};\\ \cdots;
\underbrace{i_1,\cdots, i_s}_{k}), (u=ks)$, we also do not consider the duplicate terms in
(\ref{partition-C1}).

Let us consider the following infinitesimal transformations, respectively,

(i) $M_j\longrightarrow M_j+\epsilon\dsum_{j=1}^p\dsum_{n=0}^{\infty}N^{-(n+1)}
    (n+1)t_{n+1}^{(j)}\tr(M_{j}M_{j}^{\dag})^nM_j$,

(ii) $M_j\longrightarrow M_j+\epsilon\dsum_{j,j_1,\cdots,j_r=1\atop j_a\neq
     j_{a+1}}^{p}\dsum_{n_1,\cdots,n_{r+1}=0}^{\infty}\tilde{\mathcal{N}}
     N^{-\tilde{\mathcal{N}}}t_{n_1+1,\cdots,n_{r+1}+1}^{(j,j_1,\cdots,j_r)}
     \tr(M_{j_1}M_{j_1}^{\dag})^{n_1+1}(M_{j_2}M_{j_2}^{\dag})^{n_2+1}\cdots\\
    (M_{j_r}M_{j_r}^{\dag})^{n_r+1}(M_{j}M_{j}^{\dag})^{n_{r+1}}M_j$,
where $\tilde{\mathcal{N}}=(n_1+1)+(n_2+1)+\cdots+(n_{r+1}+1)$.

From the invariance of the integral (\ref{partition-C1}), we have
\begin{eqnarray}\label{con-C0}
\hat{D}_iZ_{C}=\hat{W}_iZ_{C}, \ i=1,2,
\end{eqnarray}
where \begin{eqnarray}
\tilde{D}_1&=&\sum_{j=1}^p\sum_{n=0}^{\infty}
(n+1)N^{-(n+1)}t_{n+1}^{(j)}\frac{\partial}{\partial t_{n+1}^{(j)}},\nonumber\\
\tilde{D}_2&=&\sum_{j,j_1,\cdots,j_r=1\atop j_a\neq j_{a+1}}^{p}
\sum_{n_1,\cdots,n_{r+1}=0}^{\infty}
\tilde{\mathcal{N}}N^{-\tilde{\mathcal{N}}}t_{n_1+1,\cdots,n_{r+1}+1}^{(j,j_1,\cdots,j_r)}
\frac{\partial}{\partial t_{n_1+1,\cdots,n_r+1,n_{r+1}+1}^{(j,j_1,\cdots,j_r)}},
\end{eqnarray}
and
\begin{eqnarray}
&&\tilde{W}_1=\sum_{j=1}^p\sum_{n=0}^{\infty}
\{
(n+1)N^{-(n+1)}t_{n+1}^{(j)}2N\frac{\partial}{\partial t_n^{(j)}}(1-\delta_{n,0})
+N^{-1}t_{1}^{(j)}N^2\delta_{n,0}
\nonumber\\&&
+\sum_{a=1}^{n-1}(n+1)N^{-(n+1)}t_{n+1}^{(j)}\frac{\partial}{\partial
t_a^{(j)}}\frac{\partial}{\partial t_{n-a}^{(j)}}
+\sum_{n=0}^{\infty}\sum_{k=0}^{\infty}kN^{-k-(n+1)}t_{k}^{(j)}(n+1)t_{n+1}^{(j)}
\frac{\partial}{\partial t_{n+k}^{(j)}}
\nonumber\\&&
+\sum_{u=2}^{\infty}\sum_{\substack{i_1,\cdots,i_u\\i_1,i_u\neq
j}}\sum_{k,k_{1},\cdots,k_u=1}^{\infty}k
(n+1)N^{-(n+1)-(k+k_1+\cdots+k_u)}t_{n+1}^{(j)}t_{k,k_1,\cdots k_{u}}^{(j,i_1,\cdots,i_u)}
\frac{\partial}{\partial t_{k+n,k_1,\cdots,k_u}^{(j,i_1,\cdots,i_u)}}
\},\nonumber\\&&
\tilde{W}_2=\sum_{j,j_1,\cdots,j_r=1\atop j_a\neq j_{a+1}}^{p}\sum_{n_1,\cdots,n_{r+1}=0}^{\infty}
\tilde{\mathcal{N}}N^{-\tilde{\mathcal{N}}}
\{
t_{n_1+1,\cdots,n_{r}+1,1}^{(j,j_1,\cdots,j_r)}N\frac{\partial}{\partial
t_{n_1+1,\cdots,n_r+1}^{(j_1,\cdots,j_r)}}\delta_{n_{r+1},0}
\nonumber\\&&
+t_{n_1+1,\cdots,n_{r+1}+1}^{(j,j_1,\cdots,j_r)}
[N\frac{\partial}{\partial t_{n_1+1,\cdots,n_r+1,n_{r+1}}^{(j_1,\cdots,j_r,j)}}
+\sum_{a=1}^{n_{r+1}-1}\frac{\partial^2}
{\partial t_{n_1+1,\cdots,n_r+1,a}^{(j_1,\cdots,j_r,j)}\partial t_{n_{r+1}-a}^{(j)}}
\nonumber\\&&
+\sum_{s=2}^{r-1}\sum_{a=1}^{n_s}\frac{\partial^2}
{\partial t_{n_1+1,\cdots,n_{s-1}+1,a}^{(j_1,\cdots,j_{s-1})}
\partial t_{n_s+1-a+n_{r+1},n_{s+1}+1,n_{s+2}+1,\cdots,n_r+1}^{(j_s,j_{s+1},\cdots,j_r)}}
](1-\delta_{n_{r+1},0})
\}
\nonumber\\&&
+\sum_{j,j_1=1\atop j\neq j_{1}}^{p}\sum_{n_1,n_{2}=0}^{\infty}\tilde{\mathcal{N}}
N^{-(n_1+1+n_2+1)-1}t_{n_1+1,n_{2}+1}^{(j,j_1)}t_{1}^{(j)}
\frac{\partial}{\partial t_{n_1+1,n_2+1}^{(j_1,j)}}
\nonumber\\&&
+
\sum_{j,j_1,\cdots,j_r=1\atop j_a\neq j_{a+1}}^{p}
\sum_{n_1,\cdots,n_{r+1}=0}^{\infty}\tilde{\mathcal{N}}N^{-\tilde{\mathcal{N}}}
t_{n_1+1,\cdots,n_{r+1}+1}^{(j,j_1,\cdots,j_r)}
\{N^{-1}t_{1}^{(j)}
\frac{\partial}{\partial t_{n_1+1,\cdots,n_{r+1}+1}^{(j_1,\cdots,j_r,j)}}
\nonumber\\
&&
+\sum_{k=2}^{\infty}N^{-k}
kt_{k}^{(j)}
\frac{\partial}{\partial t_{k+n_{r+1},n_1+1,\cdots,n_r+1}^{(j_1,\cdots,j_r,j)}}
+\sum_{\substack{i\neq j\\(r\geq2)}}\sum_{k=1}^{\infty}
N^{-(1+k)}t_{1,k}^{(j,i)}
\frac{\partial}{\partial t_{n_1+1,\cdots,n_r+1,n_{r+1}+1,k}^{(j_1,\cdots,j_r,j,i)}}
\nonumber\\&&
+\sum_{\substack{i\neq j\\(r=1)}}\sum_{k=1}^{\infty}
N^{-(1+k)}t_{1,k}^{(j,i)}\frac{\partial}{\partial t_{n_1+1,n_2+1,k}^{(j_1,j,i)}}
+\sum_{\substack{i\neq j;i=j_1\\(r\geq2)}}\sum_{k=1}^{\infty}N^{-(1+k)}t_{1,k}^{(j,j_1)}
\frac{\partial}{\partial t_{n_1+1+k,n_2+1,\cdots,n_{r+1}+1}^{(j_1,\cdots,j_r,j)}}
\nonumber\\&&
+\sum_{\substack{i\neq j\\i=j_1(r=1)}}\sum_{k=1}^{\infty}
N^{-(1+k)}t_{1,k}^{(j,j_1)}\frac{\partial}{\partial t_{n_1+1+k,n_2+1}^{(j_1,j)}}
+\sum_{u=2}^{\infty}\sum_{k_1,\cdots,k_u=1}^{\infty}
[N^{-(1+k_1+\cdots+k_u)}
(
\sum_{\substack{i_1,\cdots,i_u\\i_a\neq i_{a+1}}}
t_{1,k_1,\cdots k_u}^{(j,i_1,\cdots,i_u)}
\nonumber\\&&
\cdot\frac{\partial}{\partial
t_{n_1+1,\cdots,n_r+1,n_{r+1}+1,k_1,\cdots,k_u}^{(j_1,\cdots,j_r,j,i_1,\cdots,i_u)}}
+\sum_{\substack{i_1,\cdots,i_u\\i_a\neq i_{a+1},i_u=j}}
t_{1,k_1,\cdots k_u}^{(j,i_1,\cdots,i_u)}
\frac{\partial}{\partial t_{n_1+k_u+1,\cdots,n_r+1,n_{r+1}+1,k_1,\cdots,k_{u-1}}
^{(j_1,\cdots,j_r,j,i_1,\cdots,i_{u-1})}}
)
\nonumber\\&&
+\sum_{\substack{i_1,\cdots,i_u\\i_1,i_u\neq j;i_{a}\neq i_{a+1}}}
\sum_{k=2}^{\infty}\sum_{a=1}^{k-1} k N^{-(k+k_1+\cdots+k_u)}t_{k,k_1,\cdots
k_u}^{(j,i_1,\cdots,i_u)}
\frac{\partial}{\partial t_{a,n_1+1,\cdots,n_r+1,n_{r+1}+k-a,k_1,\cdots,k_u}
^{(j,j_1,\cdots,j_r,j,i_1,\cdots,i_u)}}]
\}.
\end{eqnarray}

Then let us focus on the sum of (\ref{con-C0}), i.e.,
\begin{eqnarray}\label{constraint-C}
\tilde{D}Z_{C}=\tilde{W}Z_{C},
\end{eqnarray}
where $\tilde{D}=\tilde{D}_1+\tilde{D}_2, \tilde{W}=\tilde{W}_1+\tilde{W}_2$.

The partition function (\ref{partition-C1}) can be expanded into the grading form
$Z_{C}\{t\}=\dsum_{s=0}^{\infty}Z_{C}^{(s)}$, where
\begin{eqnarray}
&&Z_{C}^{(s)}=e^{N(\sum_{i=1}^p t_0^{(i)})}
\sum_{l=0}^{\infty}\frac{1}{l!}\sum_{\substack{L=l\\\rho=s}}
\langle\prod_{i_1=1}^{l_1}\tr(M_{1}M_{1}^{\dag})^{s_{i_1}^{(1)}}\prod_{i_2=1}^{l_2}\tr
(M_{2}M_{2}^{\dag})^{s_{i_2}^{(2)}}\cdots
\prod_{i_p=1}^{l_p}\tr(M_{p}M_{p}^{\dag})^{s_{i_p}^{(p)}}
\nonumber\\&&
\prod_{i=1}^{l_{p+1}}\tr(M_{i_1}M_{i_1}^{\dag})^{k_i^{(1)}}
(M_{i_2}M_{i_2}^{\dag})^{k_i^{(2)}}\cdots(M_{i_u}M_{i_u}^{\dag})^{k_i^{(u)}}\rangle
\cdot
\prod_{i_1=1}^{l_1}t_{s_{i_1}}^{(1)}\prod_{i_2=1}^{l_2}t_{s_{i_2}}^{(2)}\cdots
\prod_{i_p=1}^{l_p}t_{s_{i_p}}^{(p)}
\prod_{i=1}^{l_{p+1}}t_{[k_i^{(1)},\cdots,k_{i}^{(u)}]}^{[i_1,\cdots,i_u]}
\nonumber\\&&
\cdot N^{-\rho}\int d^2M_1d^2M_2\cdots d^2M_p \exp(-\sum_{i=1}^p\tr
M_iM_i^{\dag}),
\end{eqnarray}
in which~ $L=\dsum_{i=1}^{p+1}l_i$, $\rho=\dsum_{i=1}^{p+1}\rho_i$,
~$\rho_{a}=\dsum_{i_a=1}^{l_a}s_{i_a}^{(a)} \ (a=1,\cdots,p),~
\rho_{p+1}=\dsum_{i=1}^{l_{p+1}}(k_{i}^{(1)}+\cdots+k_{i}^{(u)})$,
and correlators are given by
\begin{eqnarray}
\langle\cdots\rangle=\frac{\int d^2M_1d^2M_2\cdots d^2M_p \cdots
\exp(-\sum_{i=1}^p \tr M_iM_i^{\dag})}{\int d^2M_1d^2M_2\cdots d^2M_p
\exp(-\sum_{i=1}^p \tr M_iM_i^{\dag})}.
\end{eqnarray}

The degrees of operators are defined as
$deg(t_k^{(j)})=k, deg(\frac{\partial}{\partial t_k^{(j)}})=-k,$ and
$deg(\frac{\partial}{\partial t_{k_1,k_2,\cdots,k_{u}}^{(j_1,j_2,\cdots,j_u)}})
=-(k_1+k_2+\cdots+k_{u})$, then we have $\deg (\tilde D)=0, \deg (\tilde W)=1$ \cite{2matrix}.
Since the operators $\tilde{D}$ and $\tilde{D}-\tilde{W}$ being invertible and
$\tilde{D}e^{N(\sum_{j=1}^p t_0^{(j)})}=0$, from (\ref{constraint-C}), we have
\begin{eqnarray}
\dsum_{s=1}^{\infty}Z_{C}^{(s)}
=(\tilde{D}-\tilde{W})^{-1}\tilde{W}e^{N(\sum_{j=1}^p t_0^{(j)})}
=\dsum_{k=1}^{\infty}(\tilde{D}^{-1}\tilde{W})^{k}e^{N(\sum_{j=1}^p t_0^{(j)})}.
\end{eqnarray}
Note that~$\tilde{W}$ is an homogeneous operator with degree~$1$,
and~$\tilde{D}f=deg(f)\cdot f$ for any homogeneous function~$f$.
Then the $W$-representation of complex multi-matrix model can be obtained
\begin{eqnarray}\label{wrep-zc}
Z_{C}= e^{ \tilde{W}}e^{N(\sum_{i=1}^p t_0^{(i)})}.
\end{eqnarray}

By means of the $W-$representation (\ref{wrep-zc}), we obtain
\begin{eqnarray}\label{correlator-C}
&&\langle\prod_{i_1=1}^{l_1}\tr(M_1M_1^{\dag})^{s_{i_1}^{(1)}}
\cdots\prod_{i_p=1}^{l_p}\tr(M_pM_p^{\dag})^{s_{i_p}^{(p)}}
\prod_{i=1}^{l_{p+1}}\tr(M_{i_1}M_{i_1}^{\dag})^{k_i^{(1)}}(M_{i_2}M_{i_2}^{\dag})^{k_i^{(2)}}\cdots
(M_{i_u}M_{i_u}^{\dag})^{k_i^{(u)}}
\rangle
\nonumber\\&&
=\frac{l_1!l_2!\cdots l_{p+1}!}{(m+1)!\lambda(s_1^{(1)},\cdots,s_{l_1}^{(1)})\cdots
\lambda(s_1^{(p)},\cdots,s_{l_p}^{(p)})\lambda(k_1^{(1)},\cdots,k_1^{(u)};\cdots;k_{l_{p+1}}^{(1)},
\cdots,k_{l_{p+1}}^{(u)})}
\nonumber\\&&
\sum_{\rho=1}^{m+1}\sum_{\sigma}
P^{[\sigma(s_1^{(1)}),\cdots,\sigma(s_{l_1}^{(1)})];\cdots;[\sigma(s_1^{(p)}),\cdots,\sigma(s_{l_p}^{(p)})];
[\sigma(k_1^{(1)}),\cdots,\sigma(k_1^{(u)});\cdots;\sigma(k_{l_{p+1}}^{(1)}),\cdots,\sigma(k_{l_{p+1}}^{(u)})]},
\end{eqnarray}
where $P^{[\sigma(s_1^{(1)}),\cdots,\sigma(s_{l_1}^{(1)})];\cdots;[\sigma(s_1^{(p)}),\cdots,\sigma(s_{l_p}^{(p)})];
[\sigma(k_1^{(1)}),\cdots,\sigma(k_1^{(u)});\cdots;\sigma(k_{l_{p+1}}^{(1)}),\cdots,\sigma(k_{l_{p+1}}^{(u)})]}$ is the coefficients before $N^{-\rho}\dprod_{i_1=1}^{l_1}t_{s_{i_1}}^{(1)}\dprod_{i_2=1}^{l_2}t_{s_{i_2}}^{(2)}\cdots
\dprod_{i_p=1}^{l_p}t_{s_{i_p}}^{(p)}\dprod_{i=1}^{l_{p+1}}t_{k_i^{(1)},\cdots,k_{i}^{(u)}}^{(i_1,\cdots,i_u)}
$ in $\tilde{W}^{m+1}$.

For example, we list some correlators ~$ (j=1,2,\cdots,p;j+1\leq p)$
\begin{eqnarray}\label{corr-c}
\begin{array}{ll}
\langle \tr M_jM_j^{\dag}\rangle=N^2 ,
~~~~~~~~~~~~\quad\quad\quad\quad
\langle \tr (M_jM_j^{\dag})^2\rangle=2N^3  ,
\\
\langle \tr (M_jM_j^{\dag}M_{j+1}M_{j+1}^{\dag}) \rangle=N^3 \  ,
~~~~~
\langle \tr (M_jM_j^{\dag})^3\rangle=N^2+5N^4 \ ,
\\
\langle \tr M_jM_j^{\dag}(M_{j+1}M_{j+1}^{\dag})^2\rangle=6N^4 \  ,
~~~~
\langle \tr M_jM_j^{\dag}\tr M_jM_j^{\dag}\rangle=(N^2+1)N^2 \ ,
\\
\langle \tr M_jM_j^{\dag}\tr M_{j+1}M_{j+1}^{\dag}\rangle=2 N^4,
~~~~~~
\langle \tr M_jM_j^{\dag}\tr (M_jM_j^{\dag})^2\rangle=8(N^3+N^5)\ ,
\\
\langle \tr M_jM_j^{\dag}\tr M_jM_j^{\dag}\tr M_jM_j^{\dag}\rangle=(N^2+2)(N^2+1)N^2 \ ,
\\
\langle (\tr M_jM_j^{\dag})^2\tr M_{j+1}M_{j+1}^{\dag}\rangle
=\langle \tr M_jM_j^{\dag}(\tr M_{j+1}M_{j+1}^{\dag})^2\rangle=3(N^4+N^6).
\end{array}
\end{eqnarray}

The one-matrix differential formulation of two-matrix models was derived in Ref. \cite{onedefor}.
For the complex multi-matrix model (\ref{partition-C1}), it can be expressed as the following differential reformulation:
\begin{small}
\begin{eqnarray}\label{refor}
Z_{C}\{t\}
&=&\prod_{k\geq1}\prod_{i=1}^p\sum_{a_k^{(i)}=0}^{\infty}\frac{(N^{-k}t_k^{(i)})^{a_k^{(i)}}}{a_k^{(i)}!}
\prod_{u\geq2;\atop l_1,\cdots,l_u\geq1}\prod_{i_1,\cdots,i_u=1;\atop i_a\neq i_{a+1}}^{p}
\sum_{b_{l_1,\cdots,l_u}^{(i_1,\cdots,i_u)}\geq0}
\frac{(N^{-(l_1+\cdots+l_u)}t_{l_1,\cdots,l_u}^{(i_1,\cdots,i_u)})^{b_{l_1,\cdots,l_u}^{(i_1,\cdots,i_u)}}}
{b_{l_1,\cdots,l_u}^{(i_1,\cdots,i_u)}!}
\nonumber\\&&
\sum_{i\geq0}\frac{(-1)^i}{i!}\sum_{j=1}^p\sum_{k_j=0}^{n-\sum_{j=1}^{p-1}k_j}\prod_{r=0}^{p-1}
\binom{k_{r+1}}{n-\sum_{m=0}^rk_m}\prod_{n=1}^p(\tr\frac{\partial}{\partial M_n\partial
M_n^{\dag}})^{k_n}(\tr(M_1M_1^{\dag})^k)^{a_{k}^{(1)}}\nonumber\\&&
\cdots(\tr(M_pM_p^{\dag})^k)^{a_{k}^{(p)}}(\tr(M_{i_1}M_{i_1}^{\dag})^{l_1}(M_{i_2}M_{i_2}^{\dag})^{l_2}
\cdots(M_{i_u}M_{i_u}^{\dag})^{l_u})^{b_{l_1,\cdots,l_u}^{(i_1,\cdots,i_u)}}|_{M_1=\cdots=M_p=0}
\nonumber\\
&=&[e^{-\tr(\frac{\partial^2}{\partial M_1\partial M_1^{\dag}}+\cdots
+\frac{\partial^2}{\partial M_p\partial M_p^{\dag}})}e^{V(M_1,\cdots,M_p)+\tilde{V}(M_{i_1},\cdots,M_{i_u})}]_{M_1=\cdots=M_p=0},
\end{eqnarray}
\end{small}
where the potentials $V(M_1,\cdots,M_p)=\dsum_{i=1}^p\dsum_{k=0}^{\infty}N^{-k}t_k^{(i)}\tr(M_iM_i^{\dag})^k$
and
$\tilde{V}(M_{i_1},\cdots,M_{i_u})=\dsum_{u=2}^{\infty}\\\dsum_{i_1,\cdots,i_u=1;\atop i_a\neq
i_{a+1}}^p\dsum_{l_1,\cdots,l_u=0}^{\infty}N^{-(l_1+\cdots+l_u)}t_{l_1,\cdots ,l_{u}}^{(i_1,\cdots,i_u)}
\tr(M_{i_1}M_{i_1}^{\dag})^{l_1}(M_{i_2}M_{i_2}^{\dag})^{l_2}
\cdots(M_{i_u}M_{i_u}^{\dag})^{l_u}$
have the same definition with potentials in (\ref{partition-C1}), the propagator $\tr\frac{\partial^2}{\partial M_i\partial M_i^{\dag}}=
\dsum_{i_a,j_a=1}^N\frac{\partial}{\partial M_{i_a,j_a}}\frac{\partial}{\partial M_{j_a,i_a}^{\dag}}$.

\subsection{Correspondence between the Fredkin spin chain and connected correlators of complex multi-matrix model}
The connected operator $\tr(M_{i_1}M_{i_1}^{\dag})^{k_i^{(1)}}(M_{i_2}M_{i_2}^{\dag})^{k_i^{(2)}}
\cdots(M_{i_u}M_{i_u}^{\dag})^{k_i^{(u)}}$ can be represented by polygon (red-green cycles) of the
size $2(k_i^{(1)}+\cdots+k_i^{(u)})$ \cite{1710}, i.e.,
\begin{eqnarray}\label{polygon}
\includegraphics[width=0.65\textwidth ]{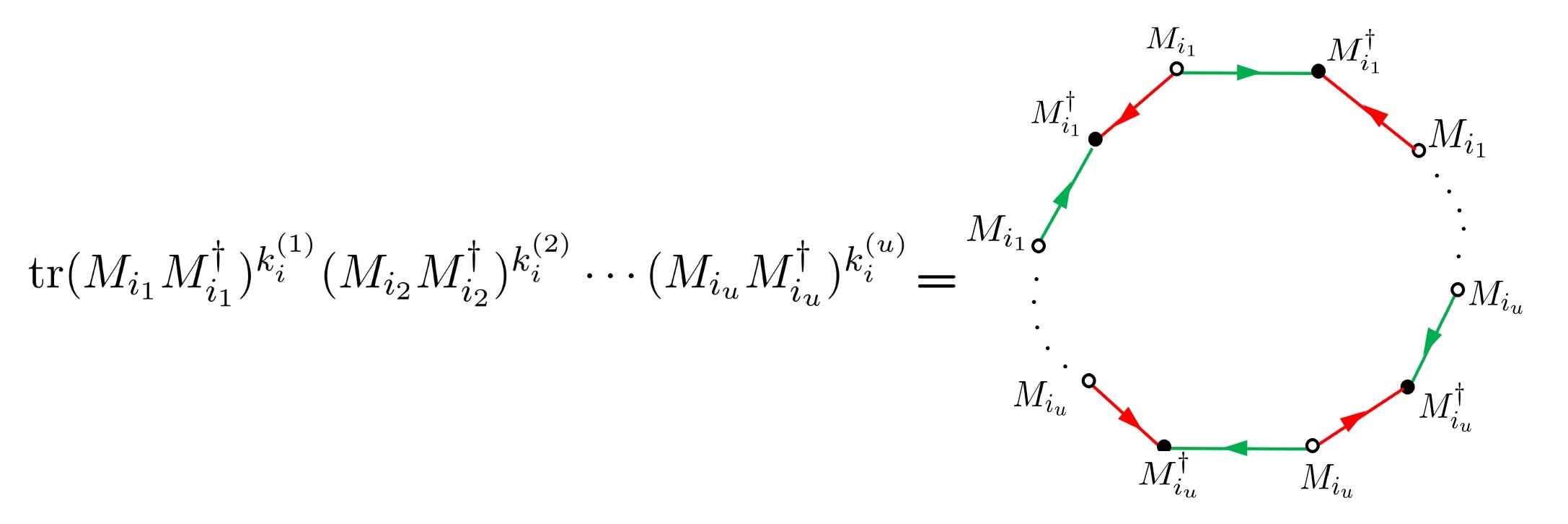},
\end{eqnarray}
where the vertices represent the matrices $M_i$, $M_j^{\dag}$, the red and green lines represent the
two indices of each matrix, and the directions of arrows depend on the choice of covariant and
contravariant indices. Then we use the thick black line (Feynman propagator) to merge two vertices
to depict the  connected correlators $\langle\tr(M_{i_1}M_{i_1}^{\dag})^{k_i^{(1)}}(M_{i_2}M_{i_2}
^{\dag})^{k_i^{(2)}}\cdots(M_{i_u}M_{i_u}^{\dag})^{k_i^{(u)}}\rangle$ graphically.
To facilitate further discussion, we give some examples of calculating correlates through Feynman
diagrams.

\begin{eqnarray}\label{corr-c-2}
\includegraphics[width=0.75\textwidth ]{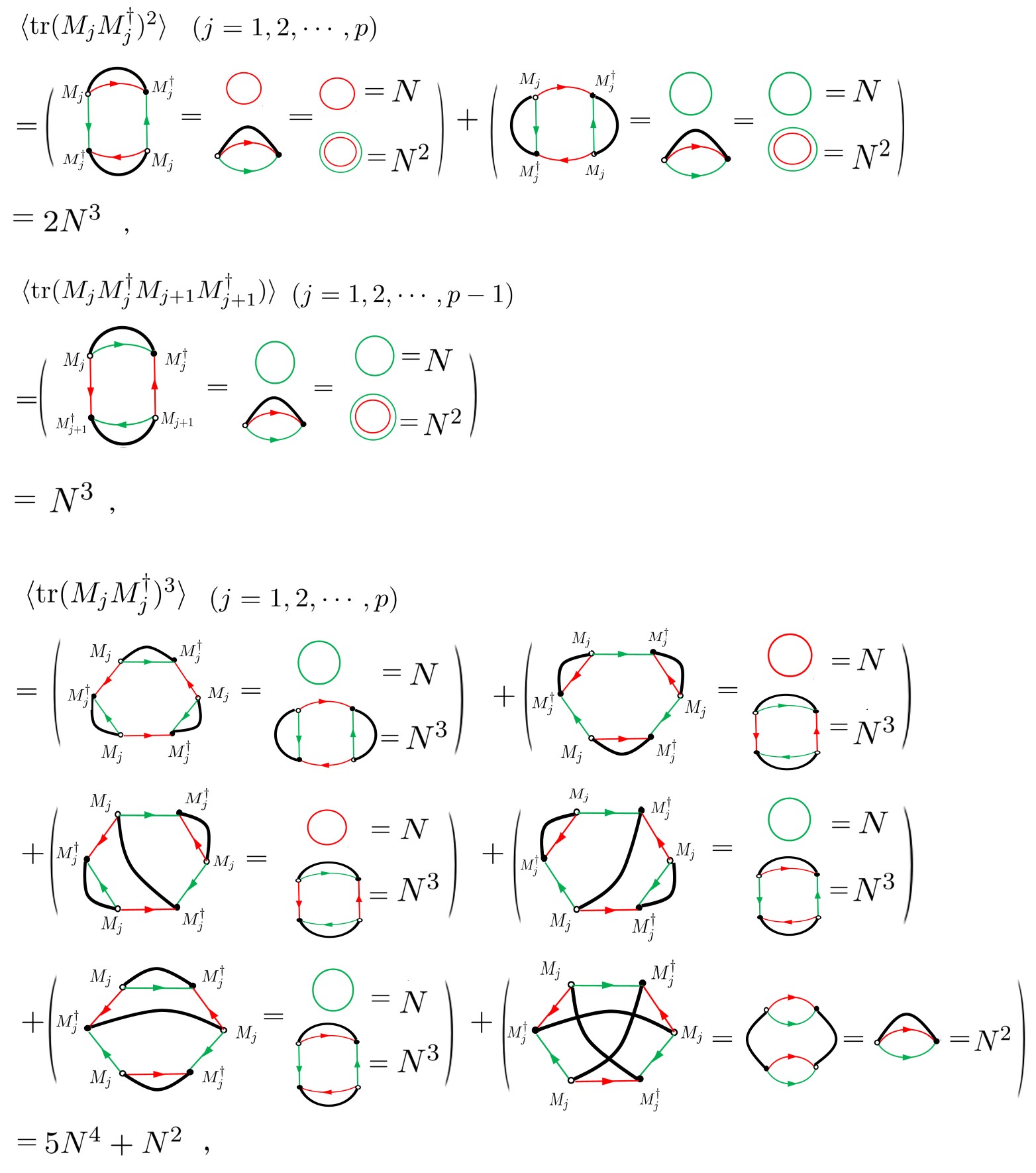}
\end{eqnarray}
where the red and green circles represent $N$.

Let us establish the correspondence between the length-$2n$ $2$-colored Dyck walks in Fredkin spin
chain and Feynman diagrams of connected correlators $\langle\tr (M_{j}M_{j}^{\dag})^{k_1} (M_{j+1}M_{j+1}^{\dag})^{k_2}(M_{j}M_{j}^{\dag})^{k_3} \\\cdots (M_{j}M_{j}^{\dag})^{k_{u-1}} (M_{j+1}M_{j+1}^{\dag})^{n-\sum_{i=1}^{u-1}k_i}\rangle$.
We make the correspondence rule as follows:

(i) For each Feynman diagram with disjoint Feynman contraction lines, we cut an arbitrary green line such that there exists a one-to-one correspondence between the cutted Feynman diagram and Dyck walk.

(ii) Since the connected correlators can give a set of Feynman diagrams with disjoint Feynman contraction lines, we cut the green line at the same position for each Feynman diagram, such that there exists a one-to-one correspondence between the cutted Feynman diagram and length-$2n$ 2-colored Dyck walk.

The cutted Feynman diagram gives a red-green chain with elements connected by disjoint black lines,
and each black line may correspond a up- and down-step. If the black line connects $M_j$
and~$M_j^{\dag}$, we paint the $j$-step with purple, otherwise with blue. We depict some examples of
the correspondence(see Figs. \ref{spin1}, \ref{spin2} and Figs. \ref{spin3-0}, \ref{spin3-1} in Appendix A).
\begin{figure}[H]
\centering
\includegraphics[width=12.5cm]{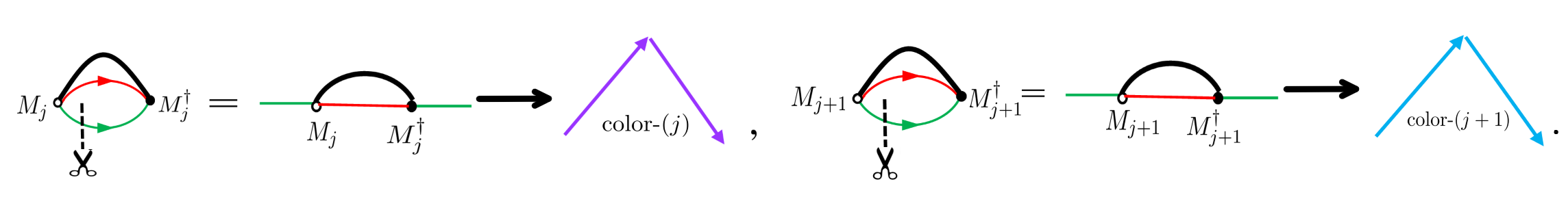}
\caption{ Correspondence between the cutted feynman diagrams and length-2 Dyck walks.}
\label{spin1}
\end{figure}
\begin{figure}[H]
\centering
\includegraphics[width=0.9\textwidth]{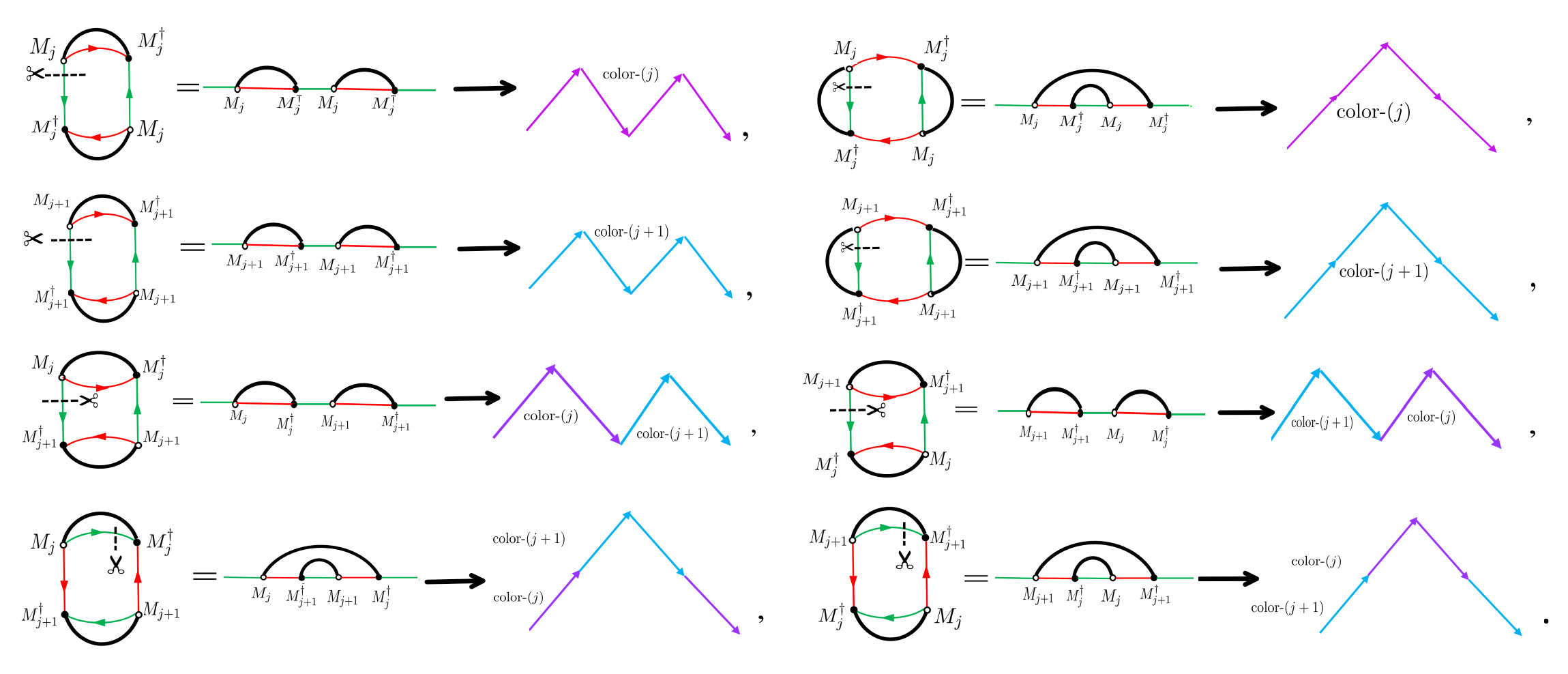}
\caption{ Correspondence between the cutted feynman diagrams and length-4 2-colored Dyck walks.}
\label{spin2}
\end{figure}

\begin{figure}[H]
\centering
\includegraphics[width=5.5cm]{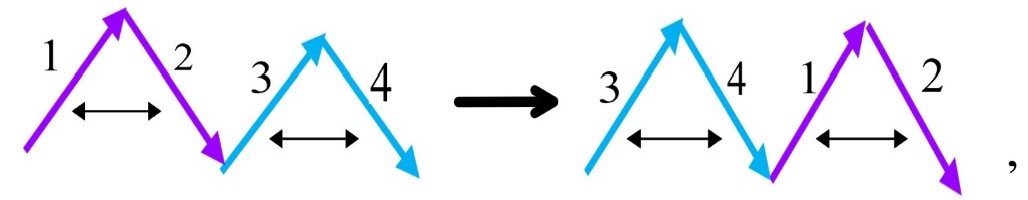}\quad
\includegraphics[width=4.85cm]{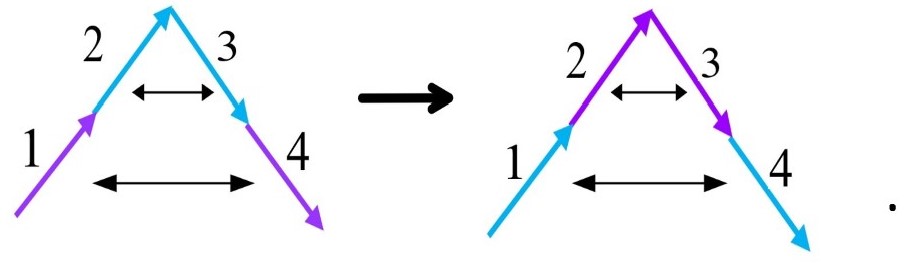}
\caption{ Two examples of moving the first two steps of Dyck walks to the back in turn.}
\label{Dyckwalks}
\end{figure}

From such correspondence, we number the steps of Dyck walks and through the similar operation with
\cite{kb2tensor}, we may obtain all Dyck walks corresponding to the Feynman diagrams of
$\langle\tr(M_jM_j^{\dag}+M_{j+1}M_{j+1}^{\dag})^n\rangle$ cutted from  green lines at same position
(see examples in Fig. \ref{Dyckwalks}).

For the number of the cutted Feynman diagrams contributing to the leading coefficient of correlator
$\langle\tr (M_{i_{1}}M_{i_{1}}^{\dag})^{k_1} (M_{i_{2}}M_{i_{2}}^{\dag})^{k_2}
\cdots (M_{i_{q}}M_{i_{q}}^{\dag})^{k_{u}} \rangle$, it is given by
$\dlim_{N\rightarrow\infty}\langle\tr(M_{j}M_{j}^{\dag})^{n}\rangle,\ (j=i_1,\cdots,i_q)$.
Let us introduce the ``cutted correlator"
$\langle\cdots\rangle_{cut}$ to calculate the number of cutted Feynman diagrams (see examples in Appendix A).
Through the correspondence rule and calculations, we have
\begin{eqnarray}
&&\langle \tr(M_{i_1}M_{i_1}^{\dag}+\cdots+M_{i_q}M_{i_q}^{\dag})^n\rangle_{cut}\nonumber\\&&
=\lim_{N\rightarrow\infty}\sum_{u=1}^{q}\binom{n}{k_1}\binom{n-k_1}{k_2}\cdots\binom{n-\sum_{i=1}^{u-1}k_i}{k_u}
\langle\tr(M_{j}M_{j}^{\dag})^{n}\rangle.
\end{eqnarray}
Then, we obtain that the number of $q$-colored Dyck walks with length-$2n$ are
\begin{eqnarray}\label{NF2nq}
N_{F,2n,q}&=&q^{n}N_{F,2n}=\frac{q^n}{n+1}\binom{2n}{n}
\nonumber\\
&=&\lim_{N\rightarrow\infty}N^{-(n+1)}\langle \tr(M_{i_1}M_{i_1}^{\dag}+\cdots+M_{i_q}M_{i_q}^{\dag})^n\rangle_{cut}
\nonumber\\
&=&\lim_{N\rightarrow\infty}\frac{q^{n}\langle\tr(M_{i_1}M_{i_1}^{\dag})^n\rangle}
{N^{n+1}}.
\end{eqnarray}
Taking $q=2$, we have the number of length-$2n$ $2$-colored Dyck walks
\begin{eqnarray}\label{NF2n2}
N_{F,2n,2}&=&2^{n}N_{F,2n}=\frac{2^n}{n+1}\binom{2n}{n}
\nonumber\\
&=&\lim_{N\rightarrow\infty}N^{-(n+1)}\langle \tr(M_jM_j^{\dag}+M_{j+1}M_{j+1}^{\dag})^n\rangle_{cut}
\nonumber\\
&=&\lim_{N\rightarrow\infty}\frac{2^n\langle \tr(M_jM_j^{\dag})^n\rangle}{N^{n+1}},
\end{eqnarray}
where $N_{F,2n}$ is the $n$-th Catalan number.

From (\ref{NF2nq}) and (\ref{NF2n2}), we have a set of correlators of the complex multi-matrix model \begin{eqnarray}\label{corre-n}
\langle\tr( M_iM_i^{\dag})^n\rangle=N_{F,2n}N^{n+1}+\circ(N^n),
\end{eqnarray}
where $i=1,\cdots,q$.

\begin{figure}[H]
\centering
\includegraphics[width=14cm]{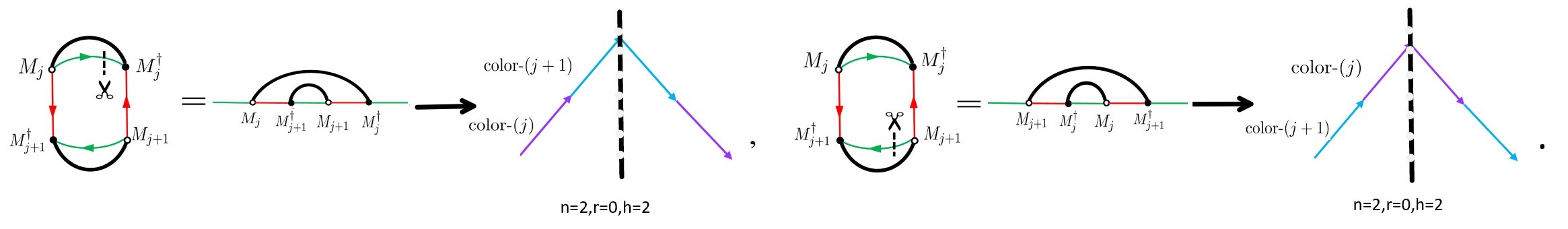}
\caption{Divide the cutted Feynman diagrams of $\langle\tr (M_{j+1}M_{j+1}^{\dag}M_jM_j^{\dag})
\rangle$ into two parts, and four colored Dyck walks of two subsystems.}
\label{cutteddiag}
\end{figure}

As done in Ref.\cite{kb2tensor}, we divide the cutted Feynman diagram into two parts by using the dotted line.
Thus the corresponding $2$-colored Dyck walks are divided into two subsystems,~$\mathcal{A}$ with $(n-r)$, and~$\mathcal{B}$
with $(n+r)$ spins, respectively.
We draw an example in the follows (see Fig. \ref{cutteddiag} ).
\begin{figure}[H]
\centering
\includegraphics[width=14cm]{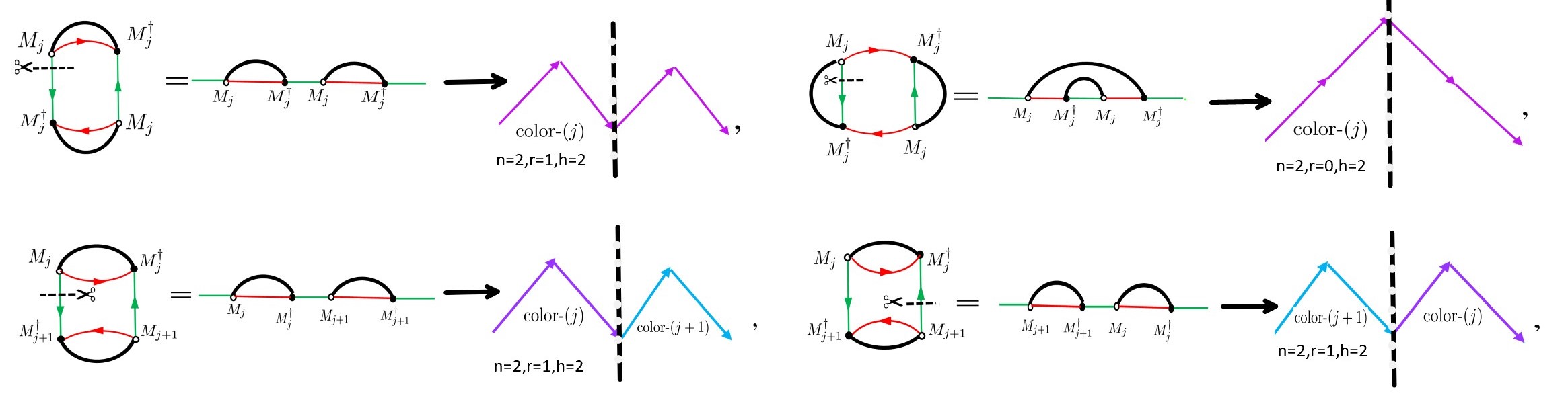}
\end{figure}
Then the number of the paths in subsystem $\mathcal{A}$ and subsystem $\mathcal{B}$, i.e. ~$N_{F,n+r,2}^{(0\rightarrow h)}$ and $N_{F,n-r,2}^{(h\rightarrow 0)}$,
are related to the number of cutted Feynman diagrams
\begin{eqnarray}
N_{F,n+r,2}^{(0\rightarrow h)}=N_{F,n-r,2}^{(h\rightarrow 0)}=2^{\gamma}N_{F,n+r}^{(h)}
=\left\{\begin{array}{cl}2^{\gamma}\frac{h+1}{\gamma+1}\binom{n+r}{\gamma}, & \gamma\in \mathbb{N},\\ 0, & otherwise.
\end{array}\right.
\end{eqnarray}

The entanglement entropy of the quantum system divided into two systems is given by \cite{Fumihiko}
\begin{eqnarray}
s_{F,A}=-\sum_{h=0}^{n-|r|}2^hp_{F,n+r,n-r,2}^{(h)}\ln p_{F,n+r,n-r,2}^{(h)},
\end{eqnarray}
where
\begin{eqnarray}\label{ph}
p_{F,n+r,n-r,2}^{(h)}=2^{-2h}\frac{N_{F,n+r,2}^{(0\rightarrow h)}N_{F,n-r,2}^{(h\rightarrow
0)}}{N_{F,2n,2}},
\end{eqnarray}
and $N_{F,2n,2}$ is determined by the correlator (\ref{corre-n}) of multi-matrix model in the large $N$ limit.

\section{Free energy and large $N$ limit in complex  multi-matrix model}
Let us consider the free energy $\mathcal{F}$ of the complex multi-matrix model (\ref{partition-C1})
\begin{small}
\begin{eqnarray}\label{freeF-C}
\mathcal{F}&=&\ln Z_{C}\{t\}\nonumber\\&=&
\sum_{s,l=1}^{\infty}\sum_{\substack{\lambda\mapsto l, \\ length(\lambda)=d,\\ \rho=s}}
\frac{(-1)^{d+1}}{l!d}S(\mu)\tilde{S}(\lambda)
C^{(n_{1,1},\cdots,n_{1,p},n_{1,p+1}),\cdots,(n_{\lambda_1,1},\cdots,n_{\lambda_1,p},n_{\lambda_1,p+1})}
\nonumber\\&&
C^{(n_{\lambda_1+1,1},\cdots,n_{\lambda_1+1,p+1}),\cdots,(n_{\lambda_2,1},\cdots,n_{\lambda_2,p},n_{\lambda_2,p+1})}
\cdots
C^{(n_{\lambda_{d-1}+1,1},\cdots,n_{\lambda_{d-1}+1,p+1}),\cdots,(n_{\lambda_d,1},\cdots,n_{\lambda_d,p},n_{\lambda_d,p+1})}
\nonumber\\&&\cdot N^{-\rho}
\prod_{i_1=1}^{l_1}t_{s_{i_1}}^{(1)}\cdots\prod_{i_p=1}^{l_p}t_{s_{i_p}}^{(p)}
\prod_{i=1}^{l_{p+1}}t_{[k_i^{(1)},\cdots,k_{i}^{(u)}]}^{[i_1,\cdots,i_u]},
\end{eqnarray}
\end{small}
where
~$C^{(n_{1,1},\cdots,n_{1,p},n_{1,p+1}),\cdots,(n_{i,1},\cdots,n_{i,p},n_{i,p+1})}
=\langle\dprod_{n_{1,1}=1}^{l_1}\tr(M_{1}M_{1}^{\dag})^{S_{n_{1,1}}^{(1)}}\cdots
\dprod_{n_{1,p}=1}^{l_p}\tr(M_{p}M_{p}^{\dag})^{S_{n_{1,p}}^{(p)}}\\
\dprod_{n_{1,p+1}=1}^{l_{p+1}}\tr(M_{i_1}M_{i_1}^{\dag})^{K_{n_{1,p+1}}^{(1)}}
\cdots(M_{i_u}M_{i_u}^{\dag})^{K_{n_{1,p+1}}^{(u)}}\cdots
\dprod_{n_{i,1}=1}^{l_1}\tr(M_{1}M_{1}^{\dag})^{S_{n_{i,1}}^{(1)}}\cdots
\dprod_{n_{i,p}=1}^{l_p}\tr(M_{p}M_{p}^{\dag})^{S_{n_{i,p}}^{(p)}}\\
\prod_{n_{i,p+1}=1}^{l_{p+1}}\tr(M_{i_1}M_{i_1}^{\dag})^{K_{n_{i,p+1}}^{(1)}}
\cdots(M_{i_u}M_{i_u}^{\dag})^{K_{n_{i,p+1}}^{(u)}}\rangle$
can be given from $(\ref{correlator-C})$,  $\rho=\dsum_{i=1}^{p+1}\rho_i$,
~$\rho_{j}=\dsum_{i_j=1}^{l_j}s_{i_j}^{(j)}$\\$(j=1,\cdots,p),~
\rho_{p+1}=\dsum_{i=1}^{l_{p+1}}(k_{i}^{(1)}+\cdots+k_{i}^{(u)})$,
and
the partitions
\begin{small}
\begin{eqnarray}\label{lambdau}
&&\lambda=(1^{n_{1,1}+\cdots+n_{1,p}+n_{1,p+1}^{(1)}+\cdots+n_{1,p+1}^{(u)}},
\cdots,d^{n_{d,1}+\cdots+n_{d,p}+n_{d,p+1}^{(1)}+\cdots+n_{d,p+1}^{(u)}})
\nonumber\\&&
\mu=(1^{v_{1,1}+\cdots+v_{1,p}+v_{1,p+1}^{(1)}+\cdots+v_{1,p+1}^{(u)}},
\cdots,d^{n_{d,1}+\cdots+n_{d,p}+n_{d,p+1}^{(1)}+\cdots+n_{d,p+1}^{(u)}}),
\end{eqnarray}
\end{small}
and
\begin{footnotesize}
\begin{eqnarray}\label{suslambda}
&&S(\mu)=\binom{\sum_i n_{i,1}+\cdots+n_{i,p}+n_{i,p+1}^{(1)}+\cdots+n_{i,p+1}^{(u)}}
{v_{1,1}+\cdots+v_{1,p}+v_{1,p+1}^{(1)}+\cdots+v_{1,p+1}^{(u)}}
\binom{\substack{\sum_i n_{i,1}+\cdots+n_{i,p}+n_{i,p+1}^{(1)}+\cdots+n_{i,p+1}^{(u)}\\
 -(v_{1,1}+\cdots+v_{1,p}+v_{1,p+1}^{(1)}+\cdots+v_{1,p+1}^{(u)})}}
 {v_{2,1}+\cdots+v_{2,p}+v_{2,p+1}^{(1)}+\cdots+v_{2,p+1}^{(u)} }
\cdots 1,
\nonumber\\&&
\tilde{S}(\lambda)=(n_{1,1}+\cdots+n_{1,p}+n_{1,p+1}^{(1)}+\cdots+n_{1,p+1}^{(u)})!
\cdots(n_{d,1}+\cdots+n_{d,p}+n_{d,p+1}^{(1)}+\cdots+n_{d,p+1}^{(u)})!.
\end{eqnarray}
\end{footnotesize}

Then in large $N$ limit, the free energy are determined by the product of connected correlators
$\langle\tr(M_{i_1}M_{i_1}^{\dag})^{K^{(1)}}
\cdots(M_{i_p}M_{i_p}^{\dag})^{K^{(p)}}\rangle$, where the operator $\tr(M_{i_1}M_{i_1}^{\dag})^{K^{(1)}}
\cdots(M_{i_p}M_{i_p}^{\dag})^{K^{(p)}}$
do not contain the terms under the $k$-transposition $(\underbrace{i_1,\cdots,i_m}_{1}; \underbrace{i_1,\cdots,i_m}_{2}; \cdots; \underbrace{i_1,\cdots, i_m}_{k}), (u=km)$, i.e.,
\begin{eqnarray}\label{F2}
\mathcal{F}&\sim&\sum_{s,l=1}^{\infty}
\sum_{\substack{\lambda\mapsto l\\length(\lambda)=d}}\frac{(-1)^{d+1}}{l!d}S(\mu)\tilde{S}(\lambda)
\langle\tr(M_{i_1}M_{i_1}^{\dag})^{K_{n_{1}}^{(1)}}(M_{i_2}M_{i_2}^{\dag})^{K_{n_{1}}^{(2)}}
\cdots(M_{i_p}M_{i_p}^{\dag})^{K_{n_{1}}^{(p)}}\rangle\cdots
\nonumber\\&&
\langle\tr(M_{i_1}M_{i_1}^{\dag})^{K_{n_{d}}^{(1)}}(M_{i_2}M_{i_2}^{\dag})^{K_{n_{d}}^{(2)}}
\cdots(M_{i_p}M_{i_p}^{\dag})^{K_{n_{d}}^{(p)}}\rangle\cdot
\prod_{i=1}^{d}t_{[K_{n_{i}}^{(1)},\cdots,K_{n_{i}}^{(p)}]}^{[i_1,\cdots,i_p]}N^{-s}
\nonumber\\&=&
\sum_{l=1}^{\infty}
\sum_{\substack{\lambda\mapsto l\\length(\lambda)=d}}\frac{(-1)^{d+1}}{l!d}S(\mu)\tilde{S}(\lambda)
\cdot \prod_{a=1}^d\Lambda_a N^d\cdot
\prod_{i=1}^{d}t_{[k_{n_{i}}^{(1)},\cdots,k_{n_{i}}^{(p)}]}^{[i_1,\cdots,i_p]}+\circ(N^{d-1}).
\end{eqnarray}
where $s=\dsum_{i=1}^dK_{n_{i}}^{(1)}+\cdots+K_{n_{i}}^{(p)}$, the coefficient of leading term is
\begin{eqnarray}\label{lambda}
\Lambda_a=N_{F,2a}\prod_{i=2}^{p}\delta_{K_{n_{a}}^{(i)},0}
+\sum_{K_{n_{a}}^{(i)}>1}\{K_{n_{a}}^{(i)}+(K_{n_{a}}^{(i)}-1)+\cdots+2\}
\end{eqnarray}
and $a=K_{n_{a}}^{(1)}+\cdots+K_{n_{a}}^{(p)}$.
It is obvious that $\Lambda_a$ is the leading coefficient of the correlator
$\langle\tr(M_{i_1}M_{i_1}^{\dag})^{K_{n_{a}}^{(1)}}(M_{i_2}M_{i_2}^{\dag})^{K_{n_{a}}^{(2)}}
\cdots(M_{i_{p-1}}M_{i_{p-1}}^{\dag})^{K_{n_{a}}^{(p-1)}}(M_{i_p}M_{i_p}^{\dag})^{K_{n_{a}}^{(p)}}\rangle$
, which can be calculated through induction.

Let us study the coefficient $\Lambda_a$ (\ref{lambda}) in terms of Feynman diagrams and count the
contributed Feynman diagrams with the help of grid diagram. For convenience, we take $p=2$ and focus
on two cases of the connected correlators with $a=3$ and $a=4$, respectively.

\begin{figure}[H]
\centering
\includegraphics[width=13cm]{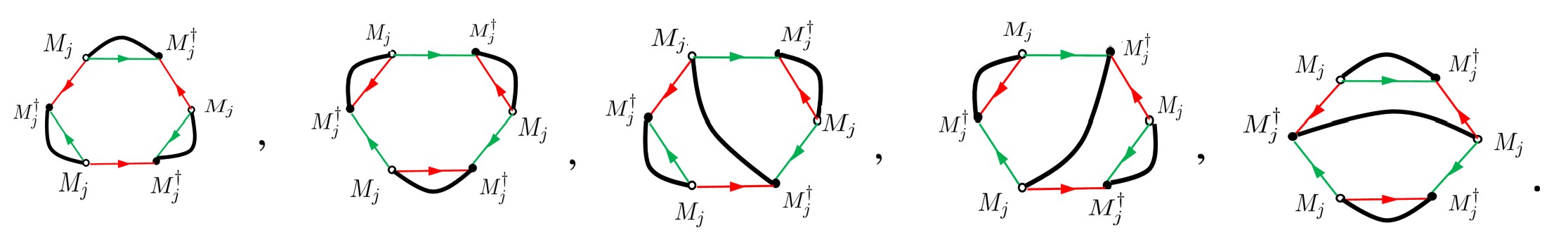}
\caption{There are five Feynman diagrams which equals to Catalan number $N_{F,6}$.}
\label{corre3num5}
\end{figure}

\begin{figure}[H]
\centering
\includegraphics[width=5.5cm]{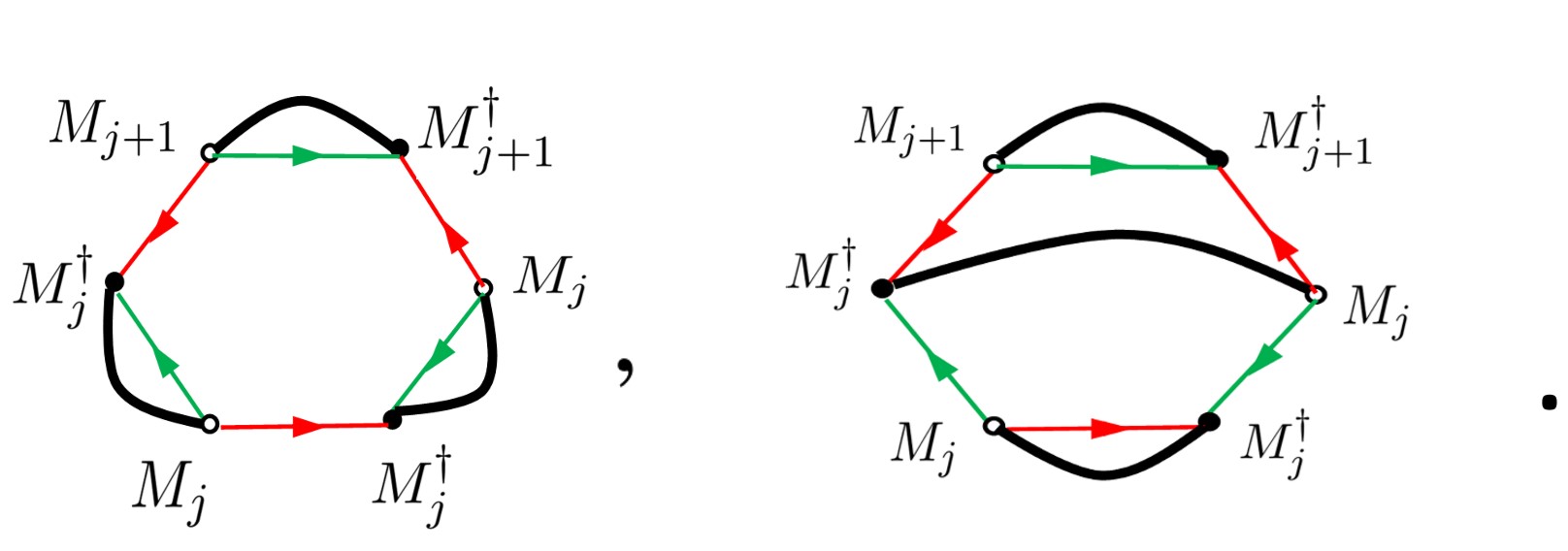}
\caption{There are two Feynman diagrams which equals to the coefficient of leading term $2N^4$.}
\label{corre3num2}
\end{figure}

For the case of $a=3$, we draw the Feynman diagrams contributing to the leading terms of
$\langle\tr(M_{j}M_{j}^{\dag})^3\rangle=5N^4+2N^2$ and $\langle\tr(M_{j}M_{j}^{\dag})^2(M_{j+1}M_{j+1}^{\dag})\rangle=2N^4$, respectively (see Figs. \ref{corre3num5} and \ref{corre3num2}).

In the follows, we present the examples of connected correlators with $a=4$.
For $\langle\tr(M_{j}M_{j}^{\dag})^4\rangle=14N^5+10N^3$, the contributing Feynman diagram is drawn as follows (see Fig. \ref{corre4num14}).
\begin{figure}[H]\label{corre4num14}
\centering
\includegraphics[width=12.5cm]{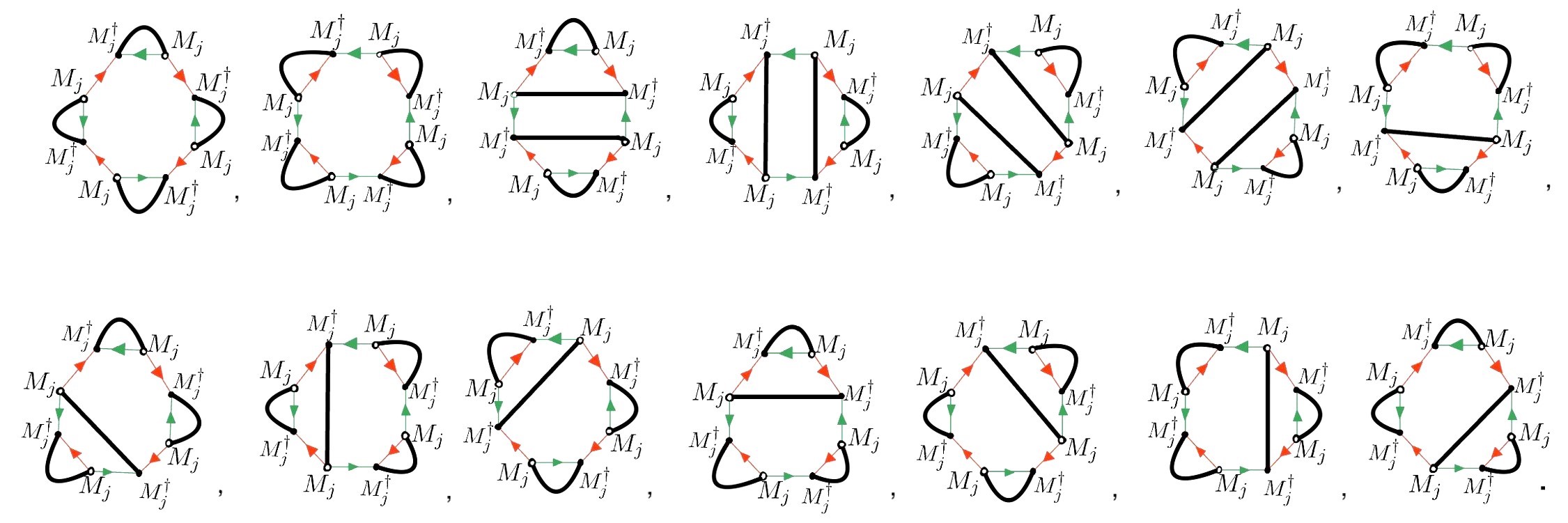}
\caption{There are fourteen Feynman diagrams which equals to $N_{F,8}$.}
\label{corre4num14}
\end{figure}

\begin{figure}[H]
\centering
\includegraphics[width=11cm]{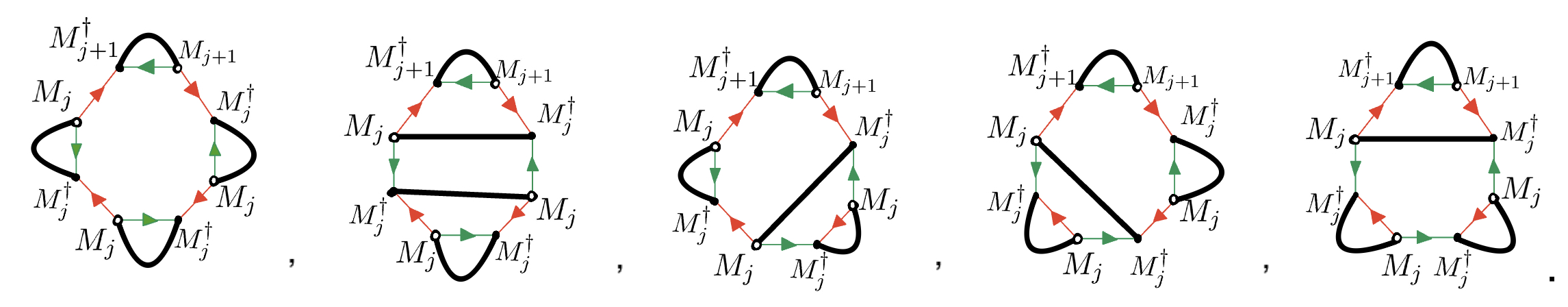}
\includegraphics[width=3.5cm]{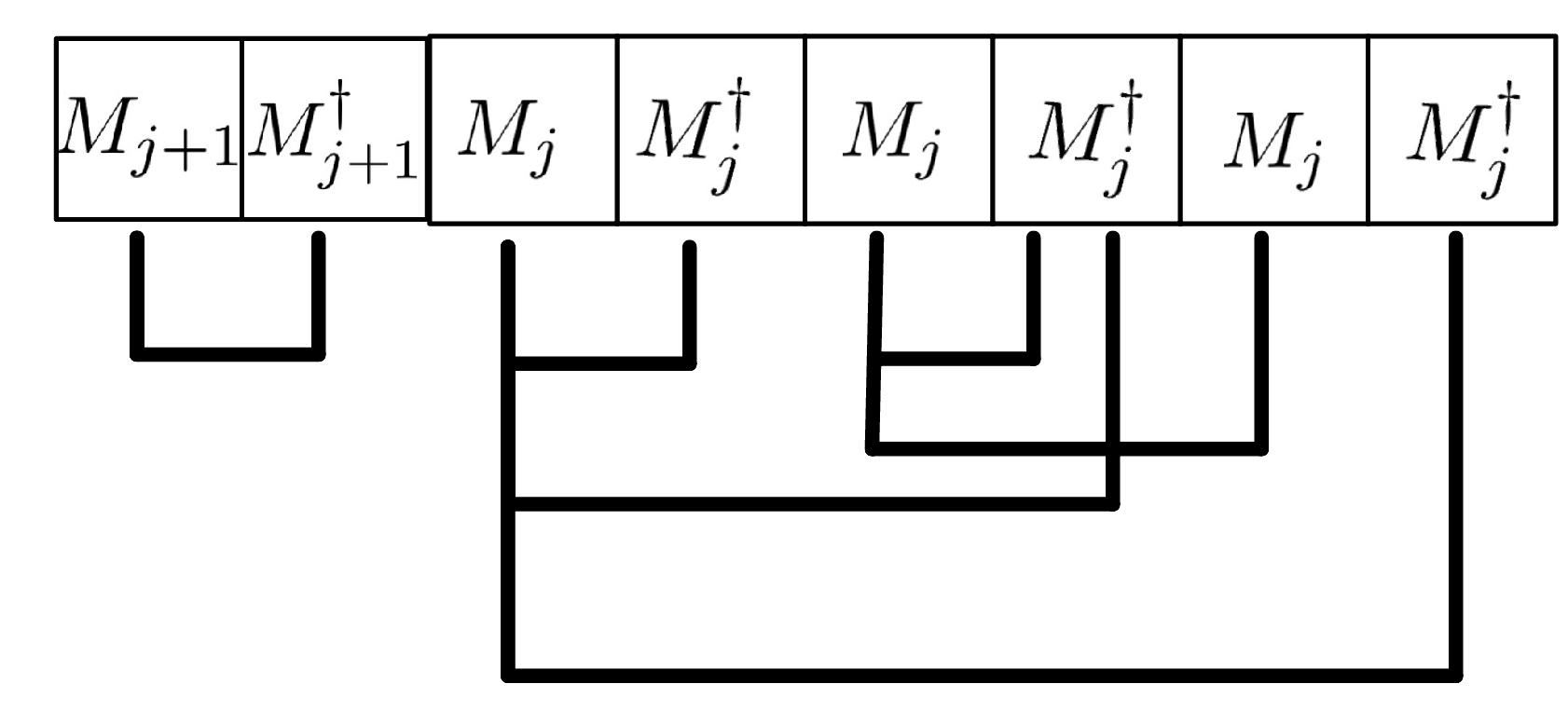}
\caption{There are five Feynman diagrams which can be calculated using the grid diagram:  $\binom{3}{1}+\binom{2}{1}$.}
\label{corre4num5}
\end{figure}

\begin{figure}[H]
\centering
\includegraphics[width=11cm]{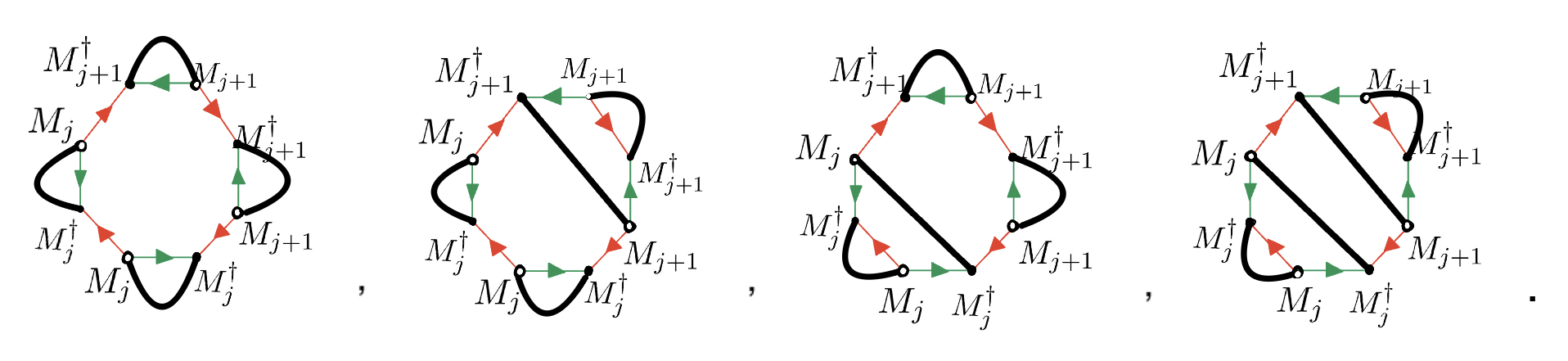}
\includegraphics[width=4cm]{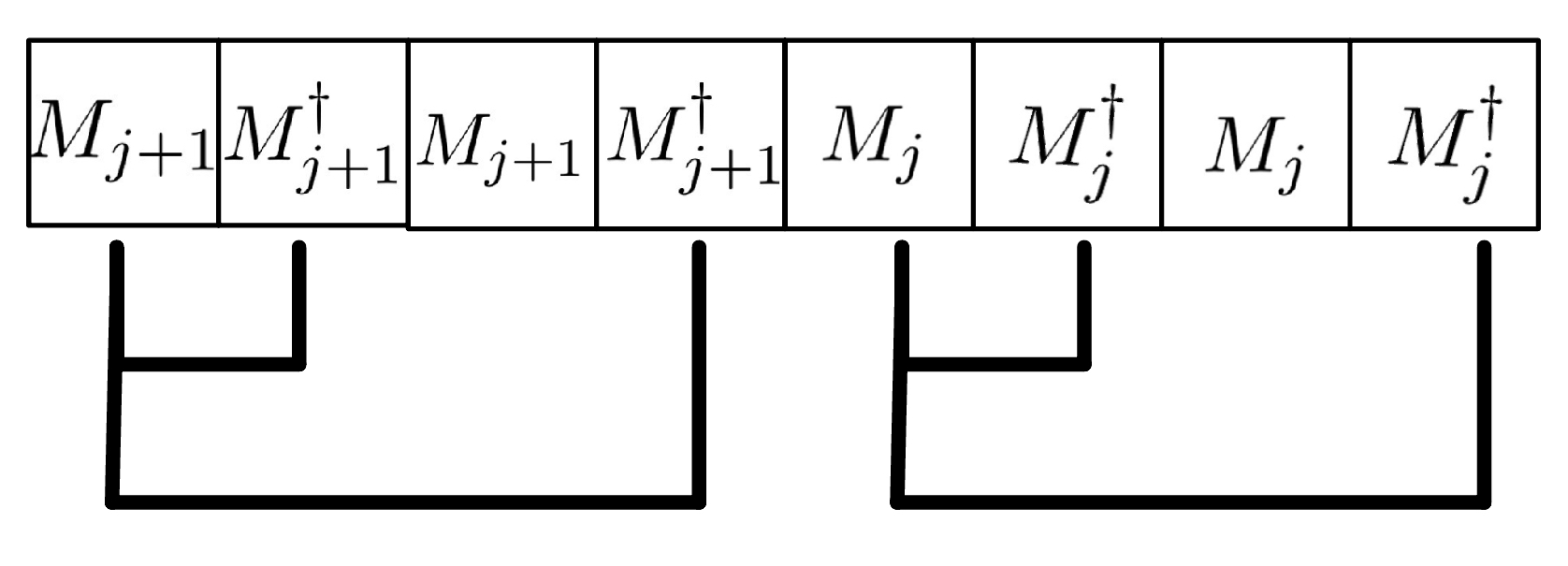}
\caption{There are four Feynman diagrams which can be calculated using the grid diagram:
$\binom{2}{1}+\binom{2}{1}$.}
\label{corre4num4}
\end{figure}

For ease of counting, we introduce thick black lines in the grid diagram to represent wick
contractions such that the number of different contractions equals to the number of Feynman
diagrams contributing to the leading term of $\mathcal{F}$.
The Feynman diagrams contribute to the leading terms of
$\langle\tr(M_{j}M_{j}^{\dag})^3(M_{j+1}M_{j+1}^{\dag})\rangle=5N^5+\circ(N^4)$
and
$\langle\tr(M_{j}M_{j}^{\dag})^2(M_{j+1}M_{j+1}^{\dag})^2\rangle=4N^5+\circ(N^4)$
are listed in Figs. \ref{corre4num5} and  \ref{corre4num4}, respectively.

\section{Conclusion}
We have constructed the complex multi-matrix model with $W$-representation and derived the compact expression of correlators. Moreover, we gave the differential formulation with respect to the matrices for this multi-matrix model. Based on the polygon description of the connected operators (\ref{polygon}),
we established the relation between the connected correlators and length-$2n$ $q$-colored Dyck walks in Fredkin spin chain. It was found that the number of length-$2n$ $q$-colored Dyck walks can be given by the connected correlators (\ref{corre-n}) in large $N$ limit. In addition, we found that the denominator of the entanglement entropy of Fredkin spin chain can be given by the connected correlator in the large $N$ limit, and the numerator can be explained by the Feynman diagrams of the connected correlators. Finally, We analyzed the free energy of complex multi-matrix model and gave its expression by the compact expressions of correlators (\ref{correlator-C}).
Furthermore, we discussed the free energy in the large $N$ limit. It was noted that the leading
coefficient of free energy is determined by the connected correlators. For further research, it
would be interesting to study the case of $\beta$-deformed multi-matrix models.

\setcounter{equation}{0}
\renewcommand\theequation{A.\arabic{equation}}
\begin{appendices}
\section{Correspondence between
the cutted Feynman diagrams and length-6 2-colored Dyck walks}
\begin{figure}[H]
\centering
\includegraphics[width=14cm]{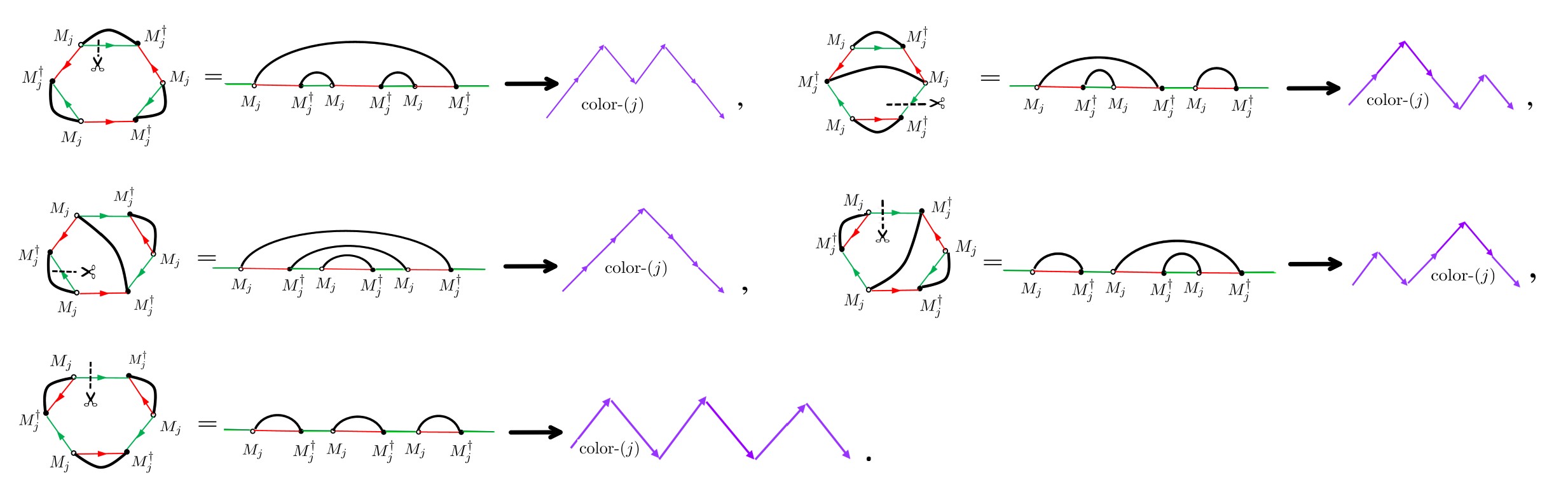}
\caption{Correspondence between $\langle\tr(M_jM_j^{\dag})^3\rangle_{cut}$  and length-$6$ $2$-colored Dyck walks.}
\label{spin3-0}
\end{figure}

\begin{figure}[H]
\centering
\includegraphics[width=14cm]{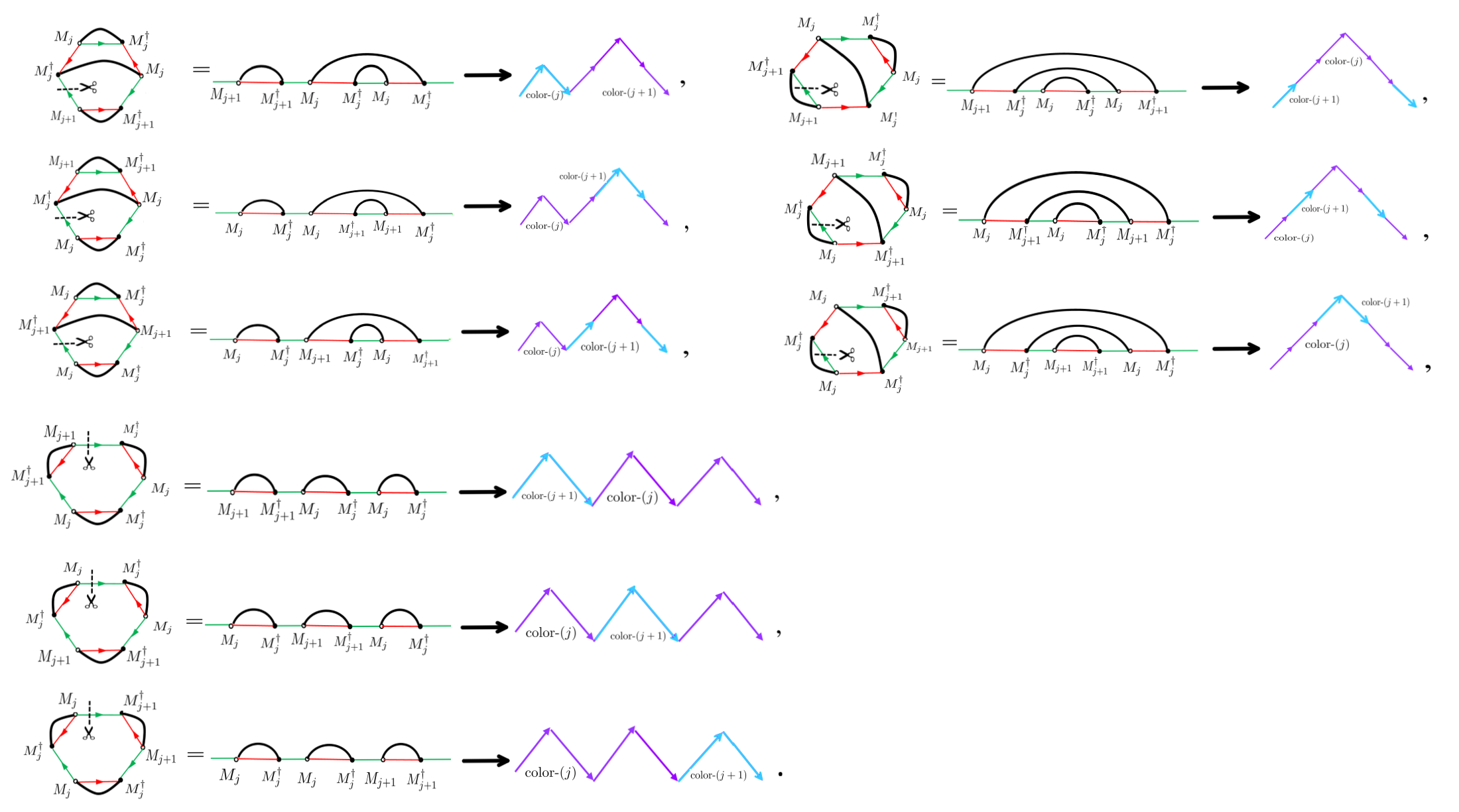}
\caption{ Correspondence between $\langle\tr(M_jM_j^{\dag})^2(M_{j+1}M_{j+1}^{\dag})\rangle_{cut}$ and length-$6$ $2$-colored Dyck walks.}
\label{spin3-1}
\end{figure}

We give an example to calculate the number of cutted Feynman diagrams,
\begin{eqnarray}
&&\langle\tr(M_jM_j^{\dag}+M_{j+1}M_{j+1}^{\dag})^3\rangle_{cut}
\nonumber\\&&=\langle\tr(M_jM_j^{\dag})^3\rangle_{cut}+\langle\tr(M_jM_j^{\dag})^2(M_{j+1}M_{j+1}^{\dag})\rangle_{cut}
+\langle\tr(M_jM_j^{\dag})(M_{j+1}M_{j+1}^{\dag})^2\rangle_{cut}
\nonumber\\&&+\langle\tr(M_{j+1}M_{j+1}^{\dag})^3\rangle_{cut}
\nonumber\\&&=(\sum_{k=0}^3\binom{3}{k})\cdot\langle\tr(M_jM_j^{\dag})^3\rangle=2^3\cdot(5N^4+N^2).
\end{eqnarray}
The cutted Feynman diagrams are drawn in Figs. \ref{spin3-0} and \ref{spin3-1}.
\begin{figure}[H]
\centering
\includegraphics[width=14cm]{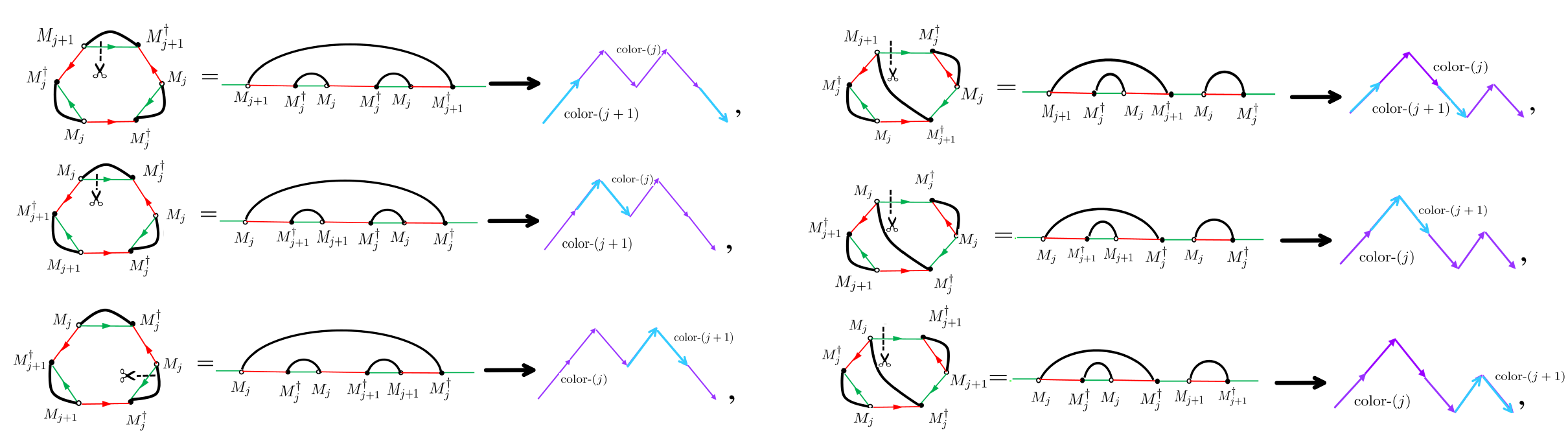}
\end{figure}
Due to the symmetry between $\langle\tr(M_jM_j^{\dag})^2(M_{j+1}M_{j+1}^{\dag})\rangle_{cut}$
and $\langle\tr(M_jM_j^{\dag})(M_{j+1}M_{j+1}^{\dag})^2\rangle_{cut}$, we do not draw the
other fifteen symmetric cutted Feynman diagrams here.
\end{appendices}

\section *{Acknowledgments}

This work is supported by the National Natural Science Foundation
of China (No. 12375004).


\end{document}